\documentclass[final,5p,times,twocolumn]{elsarticle}

\usepackage{graphicx}

\usepackage{amsmath}
\usepackage{amssymb}
\usepackage{bm}

\biboptions{comma,square}

\journal{Computer Physics Communications}

\begin{document}

\begin{frontmatter}



\title{Very-high-precision solutions of a class of Schr\"odinger equations}


\author[imf]{Asif Mushtaq}
\ead{Asif.Mushtaq@math.ntnu.no}

\author[ify]{Amna Noreen}
\ead{Amna.Noreen@ntnu.no}

\author[ify,nordita]{K{\aa}re Olaussen}
\ead{Kare.Olaussen@ntnu.no}

\author[ify]{Ingjald {\O}verb{\o}}
\ead{Ingjald.Overbo@ntnu.no}

\address[ify]{Institutt for fysikk, NTNU}

\address[imf]{Institutt for matematiske fag, NTNU}

\address[nordita]{NORDITA, Stockholm}

\begin{abstract}
We investigate a method to solve a class of
Schr{\"o}dinger equation eigenvalue problems numerically
to very high precision $P$ (from thousands to a million of decimals).
The memory requirement, and the number of high precision algebraic operations,
of the method scale essentially linearly with $P$ when
only eigenvalues are computed. However, since the 
algorithms for multiplying high precision numbers scale at
a rate between $P^{1.6}$ and $P\,\log P\,\log\log P$,
the time requirement of our method increases somewhat faster than $P^2$.
\end{abstract}


\end{frontmatter}



\section{Introduction}

The one-dimensional anharmonic oscillator
have been subject to much investigation since the seminal
works by Bender and Wu \cite{BenderWu, BenderWuII} on the
behaviour of its perturbation expansion. The motivation
has often been to extract features and understanding that
can be generalized to more interesting situations, like
quantum field theories in higher space-time dimensions,
or to test new approximation methods.

In this note we report briefly from our work on
a class of one-dimensional quantum mechanical
systems which include the one mentioned above,
i.e., systems which are modelled
by Schr{\"o}dinger equations of the type
\begin{equation}
   -s^2\psi''(x) + 
   \left(x^{2M}+\sum_{m=0}^{M-1}\,v_m\,x^{2m}\right)\,\psi(x) =  \varepsilon\,\psi(x),
   \label{TypeOfEquations}
\end{equation}
for some finite (small) $M$, and real coefficients $s$ and $v_m$. Our
focus has been on the possible accuracy to which the
eigenvalues and eigenfunctions can be found within available
computational resources (memory and CPU cycles). The algorithm we have
implemented has modest memory demands;
the required memory scales asymptotically
with $P$ like $M P$, where $P$ is the desired precision of eigenvalues or
eigenfunctions in decimal digits. The number of required algebraic
operations (involving high-precision numbers) generally also seems to
grow asymptotically with $P$ like $M P$. However, in some cases
there is a large offset which makes it computationally very expensive
to obtain $P$ to even a few digits. Further, the time required per
high-precision algebraic operation (i.e. multiplication or division)
increases somewhat faster than linearely with $P$. The high-precision
numerical library (CLN~\cite{CLN}, built
on GMP~\cite{GMP}) we have used has not been parallelized.
Thus, our algorithm is mostly constrained by wall-clock time.

Due to space constraints we can in the remainder
of this note only present examples of our
results (sections \ref{Example1}--\ref{Example3}) and a brief
decription of the requirements for obtaining a desired precision
(section \ref{ObtainablePrecision}). Solving the
differential equations (\ref{TypeOfEquations})
numerically to very high precision is the most
simple and straightforward part of our work
(section \ref{Method}); the analysis of inevitable
loss of accuracy might be more interesting
(section \ref{AccuracyLoss}).
The behaviour of our numerical algorithm
is illustrated in section \ref{NumericalObservations}.

A more complete description will be given elsewhere \cite{AAKI}. 

\section{Ground state energy to one million decimals}
\label{Example1}

As our first {\em proof-of-method\/} we considered the
ground state of the pure anharmonic oscillator,
\begin{equation}
     -\psi''(x) + x^4\,\psi(x) = \varepsilon\, \psi(x),
     \label{Anharmonic_oscillator}
\end{equation}
and computed its ground state energy to $1\,000\,000^+$ decimals.
The result, obtained after about 20 days of computing, is
\begin{align}
      &\bm{\downarrow}\text{Decimal number 1}\nonumber\\
      \varepsilon_0 =
          1&.060\,362\,090\,484\,182\,899\,647\,046\,016\,692\,663\backslash\nonumber\\
          &\,545\,515\,208\,728\,528\,977\,933\,216\,245\,241\,695\backslash\nonumber\\
          &\,943\,563\,044\,344\,421\,126\,896\,299\,134\,671\,703\backslash\nonumber\\
           &\bm{\downarrow}\text{Decimal 1\;000}\nonumber\\
           \ldots&\phantom{.}304\,916\,644\,281\,633\,946\,163\,324\,287\,004\,261\backslash\nonumber\\
           &\bm{\downarrow}\text{Decimal 10\;000}\\
           \ldots&\phantom{.}578\,044\,164\,777\,855\,042\,412\,917\,
                       855\,188\,328\backslash\nonumber\\
           &\bm{\downarrow}\text{Decimal 100\;000}\nonumber\\
           \ldots&\phantom{.}857\,326\,052\,850\,064\,563\,492\,099\,
                       229\,730\,278\backslash\nonumber\\
           &\bm{\downarrow}\text{Decimal 1\;000\;000}\nonumber\\
           \ldots&\phantom{.}820\,139\,466\,721\,621\,064\,477\,821\,481\,635\,914\ldots\nonumber
 \end{align}
The full sequence is available on request (for people seeking diversion from
investigating the numerical patterns of $\pi$).

\section{Excited states compared to WKB results}
\label{Example2}

The result of the previous section may be somewhat unsatisfactory
to sceptical readers, since there is to our knowledge no similar
results to compare against. However, it is only slightly more
challenging to treat excited states. To be fair and square we
have computed eigenvalue number $n=50\,000$ of
equation~(\ref{Anharmonic_oscillator}) to $50\,000^+$ decimals. For
such values of $n$ the WKB approximation should be resonably good.
Our result is
\begin{align}
      &\bm{\downarrow}\text{Decimal number 1}\nonumber\\
      \varepsilon_{50\,000} = 4\,024\,985&.730\,438\,698\,704\,313\,888\,104\,230\,563\backslash\nonumber\\
                                      &\phantom{.}241\,821\,769\,405\,166\,607\,313\,872\,288\backslash\nonumber\\
                                      &\phantom{.}953\,655\,475\,876\,981\,078\,813\,733\,788\backslash\nonumber\\
      &\bm{\downarrow}\text{Decimal 50\;000}\nonumber\\
      \ldots&\phantom{.}545\,947\,155\,500\,441\,209\ldots
\end{align}
In comparison, the $12^{\text{th}}$ order WKB approximation
computed by Bender {\em et. al.\/} \cite{Bender_etal} gives
\begin{align}
\varepsilon^{\text{WKB-12}}_{50\,000} = 
4\,024\,985&.730\,438\,698\,704\,313\,888\,104\,230\,563\backslash\nonumber\\
           &\phantom{.}241\,821\,769\,405\,166\,607\,313\,872\,288\backslash\nonumber\\
           &\phantom{.}953\,657\ldots
\end{align}
I.e., the relative accuracy of the WKB approximation is
\begin{align}
   \frac{\varepsilon^{\text{WKB-12}}_{50\,000} -\varepsilon_{50\,000}}{\varepsilon_{50\,000}} = 5.163\ldots\times10^{-67}.
\end{align}
From observation of the behaviour of the WKB series we find this accuracy
to be as expected for the $12^{\text{th}}$ order approximation at this value of $n$.
Some of us plan to return to a more detailed analysis of the behaviour
of the WKB approximation, which may nowadays be extended easily to much
higher orders.

\section{Brute force calculation of double-well level-splitting}
\label{Example3}

Another well analysed situation where we may stress-test our method
is the calculation of the level splitting between the lowest even and
odd parity eigenstates of the double-well potential,
\begin{equation}
     -s^2\psi''(x) + \left(x^2-1\right)^2\psi(x) = \varepsilon\, \psi(x),
   \label{DoubleWellPotential}
\end{equation}
for small $s$. The lowest even, $\varepsilon^{(+)}_0$, and
odd, $\varepsilon^{(-)}_0$, parity states are split by an
exponentially small amount
$\Delta\varepsilon_0 \equiv \varepsilon_0^{(-)}-\varepsilon_0^{(+)}$.
Asymptotically as $s \to 0^+$,
\begin{equation}
   \Delta\varepsilon_0 \sim \Delta\varepsilon^{\text{Z-J}}_0 =
   16\sqrt{\frac{2 s}{\pi}}\,\text{e}^{-4/3s}\,\text{e}^{L(s)},
  \label{ZinnJustinFormula}
\end{equation}
where $L(s)=-\left(\frac{71}{96}s+\cdots\right)$ is given
to order $s^{10}$ by Zinn-Justin\footnote{We use a different normalization:
$s=8g$ and $\varepsilon=32gE$, where $g$ and $E$ are the parameters
in \cite{ZinnJustin}.} \cite{ZinnJustin}.
We have made independent calculations of $\varepsilon_0^{(\pm)}$
to $30\,000^+$ digits accuracy for $s=\frac{1}{50\,000}$:
\begin{align}
      &\phantom{=0.000\,0}\text{Decimal 28\,954}\bm{\downarrow}\nonumber\\
     \varepsilon^{(-)}_0 &=0.000\,039\,999\,799\ldots990\,905\,404\ldots
     \\
     \varepsilon^{(+)}_0 &=0.000\,039\,999\,799\ldots984\,723\,697\ldots
\end{align}
Equation~(\ref{ZinnJustinFormula}) agrees with the difference to the expected order,
   \begin{align}
     \frac{\Delta\varepsilon_0^{\text{Z-J}}-\Delta\varepsilon_0}{\Delta\varepsilon_0}&=
     1.649\ldots\times 10^{-48} \approx 8\,052\,s^{11}.
   \end{align}
The right hand side is of the magnitude expected for the next term in $L(s)$.

We hope the three examples above have convinced the reader that it is possible
to solve the eigenvalue problems (\ref{TypeOfEquations}) to very high precision.
{\em How\/} precise will of course depend on the parameters and which eigenstate we
want to investigate.

\section{How to achieve a desired precision}
\label{ObtainablePrecision}

The eigenvalue condition for equation (\ref{TypeOfEquations})
is assumed to be that $\psi(x)\to 0$ as $x\to\pm\infty$.
We are unable to impose this exactly in our numerical
algorithm. However, there is an equivalent
Robin boundary condition which can be imposed
at some finite (large) $x$,
\begin{align}
   -s \frac{\psi'(x)}{\psi(x)} = R(x) = x^{M} + \cdots \approx\infty.
   \label{RobinApprox}
\end{align}
We don't know $R(x)$ exactly, but there is for any desired precision $P$
a finite value of $x$ such that an approximate $R(x)$ is sufficient.
The required value of $x$ can be estimated by asymptotic analysis of
equation~(\ref{TypeOfEquations}) as $x\to\infty$ (or a WKB approximation
to include estimates of constant prefactors which cannot be found
by asymptotic analysis alone). A first estimate is that
one should choose $x$ so that
\begin{equation}
    \exp\left(-\frac{2}{s}\int_{x_0}^x \sqrt{V(y)-\varepsilon}\; \text{d}y \right)  
    \equiv 10^{-P_{\text{est}}(x)} \lesssim 10^{-P},
    \label{P_Estimate}
\end{equation}
if one wants to compute the eigenvalue to $P$ decimals precision.
Here $x_0$ is the largest turning point, and to simplify we have
assumed that the Robin boundary condition is replaced
by a Diriclet one, $R(x)=\infty$.

Equation~(\ref{P_Estimate}) is an {\em a priori\/} estimate, which we have tested
by choosing a very large $x$ to obtain a very accurate eigenvalue (so that
it may be considered exact), and used this to observe the obtainable
precision for lower values of $x$.  The obtainable precision
at a given $x$ is defined as
\begin{equation}
   P(x) \equiv \lg \left| \varepsilon(x)-\varepsilon \right|,
\end{equation}
where $\varepsilon(x)$ is the eigenvalue found numerically
by use of equation~(\ref{RobinApprox}).
Figure 1 displays the difference between $P_{\text{est}}(x)$ and $P(x)$ in some cases.
Note that the difference between the estimated $P_{\text{est}}(x)$ and obtainable
$P(x)$ precision varies very little with $x$, and hence can be found numerically
from fairly low-precision calculations. The difference probably occurs because we
have neglected a slowly varying prefactor in (\ref{P_Estimate}).

\hspace{-2.7em}
\begingroup
  \makeatletter
  \providecommand\color[2][]{%
    \GenericError{(gnuplot) \space\space\space\@spaces}{%
      Package color not loaded in conjunction with
      terminal option `colourtext'%
    }{See the gnuplot documentation for explanation.%
    }{Either use 'blacktext' in gnuplot or load the package
      color.sty in LaTeX.}%
    \renewcommand\color[2][]{}%
  }%
  \providecommand\includegraphics[2][]{%
    \GenericError{(gnuplot) \space\space\space\@spaces}{%
      Package graphicx or graphics not loaded%
    }{See the gnuplot documentation for explanation.%
    }{The gnuplot epslatex terminal needs graphicx.sty or graphics.sty.}%
    \renewcommand\includegraphics[2][]{}%
  }%
  \providecommand\rotatebox[2]{#2}%
  \@ifundefined{ifGPcolor}{%
    \newif\ifGPcolor
    \GPcolorfalse
  }{}%
  \@ifundefined{ifGPblacktext}{%
    \newif\ifGPblacktext
    \GPblacktexttrue
  }{}%
  \let\gplgaddtomacro\g@addto@macro
  \gdef\gplbacktext{}%
  \gdef\gplfronttext{}%
  \makeatother
  \ifGPblacktext
    \def\colorrgb#1{}%
    \def\colorgray#1{}%
  \else
    \ifGPcolor
      \def\colorrgb#1{\color[rgb]{#1}}%
      \def\colorgray#1{\color[gray]{#1}}%
      \expandafter\def\csname LTw\endcsname{\color{white}}%
      \expandafter\def\csname LTb\endcsname{\color{black}}%
      \expandafter\def\csname LTa\endcsname{\color{black}}%
      \expandafter\def\csname LT0\endcsname{\color[rgb]{1,0,0}}%
      \expandafter\def\csname LT1\endcsname{\color[rgb]{0,1,0}}%
      \expandafter\def\csname LT2\endcsname{\color[rgb]{0,0,1}}%
      \expandafter\def\csname LT3\endcsname{\color[rgb]{1,0,1}}%
      \expandafter\def\csname LT4\endcsname{\color[rgb]{0,1,1}}%
      \expandafter\def\csname LT5\endcsname{\color[rgb]{1,1,0}}%
      \expandafter\def\csname LT6\endcsname{\color[rgb]{0,0,0}}%
      \expandafter\def\csname LT7\endcsname{\color[rgb]{1,0.3,0}}%
      \expandafter\def\csname LT8\endcsname{\color[rgb]{0.5,0.5,0.5}}%
    \else
      \def\colorrgb#1{\color{black}}%
      \def\colorgray#1{\color[gray]{#1}}%
      \expandafter\def\csname LTw\endcsname{\color{white}}%
      \expandafter\def\csname LTb\endcsname{\color{black}}%
      \expandafter\def\csname LTa\endcsname{\color{black}}%
      \expandafter\def\csname LT0\endcsname{\color{black}}%
      \expandafter\def\csname LT1\endcsname{\color{black}}%
      \expandafter\def\csname LT2\endcsname{\color{black}}%
      \expandafter\def\csname LT3\endcsname{\color{black}}%
      \expandafter\def\csname LT4\endcsname{\color{black}}%
      \expandafter\def\csname LT5\endcsname{\color{black}}%
      \expandafter\def\csname LT6\endcsname{\color{black}}%
      \expandafter\def\csname LT7\endcsname{\color{black}}%
      \expandafter\def\csname LT8\endcsname{\color{black}}%
    \fi
  \fi
  \setlength{\unitlength}{0.0500bp}%
  \begin{picture}(5040.00,3528.00)%
    \gplgaddtomacro\gplbacktext{%
      \csname LTb\endcsname%
      \put(990,843){\makebox(0,0)[r]{\strut{}$-4$}}%
      \put(990,2457){\makebox(0,0)[r]{\strut{}$4$}}%
      \put(990,3264){\makebox(0,0)[r]{\strut{}$8$}}%
      \put(990,1650){\makebox(0,0)[r]{\strut{} 0}}%
      \put(2243,220){\makebox(0,0){\strut{}$1\times10^4$}}%
      \put(3365,220){\makebox(0,0){\strut{}$2\times10^4$}}%
      \put(4486,220){\makebox(0,0){\strut{}$3\times10^4$}}%
      \put(1122,220){\makebox(0,0){\strut{} 0}}%
      \put(1804,2952){\rotatebox{90}{\makebox(0,0){\strut{}\rotatebox{-90}{$P(x)-P_{\text{est}}(x)$}}}}%
      \put(4302,770){\makebox(0,0){\strut{}$P_{\text{est}}(x)$}}%
    }%
    \gplgaddtomacro\gplfronttext{%
      \csname LTb\endcsname%
      \put(3723,3069){\makebox(0,0)[r]{\strut{}$\varepsilon_{0\phantom{1\,000}}$}}%
      \csname LTb\endcsname%
      \put(3723,2805){\makebox(0,0)[r]{\strut{}$\varepsilon_{10\,000}$}}%
      \csname LTb\endcsname%
      \put(3723,2541){\makebox(0,0)[r]{\strut{}$\varepsilon^{\text{WW}}_{0\phantom{1\,000}}$}}%
    }%
    \gplbacktext
    \put(0,0){\includegraphics{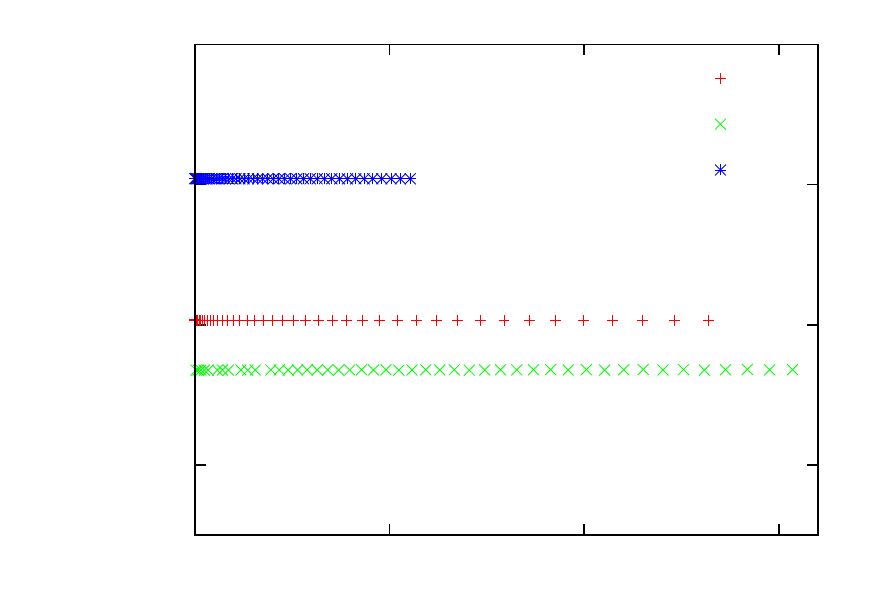}}%
    \gplfronttext
  \end{picture}%
\endgroup

\vspace{-2ex}
\begin{center}
\begin{minipage}{0.9\linewidth}{{\bf Figure 1:}
The obtainable precision $P(x)$ for various eigenvalues $\varepsilon$
with a Diriclet boundary condition
(\ref{RobinApprox}) imposed at $x$ instead of the exact condition.
The cases plotted are (i) for the lowest eigenvalue $\varepsilon_0$
of equation~(\ref{Anharmonic_oscillator}), for which
$P_{\text{est}}(x) \approx {2} x^3/{3\ln 10}$, 
(ii) for the $10\,000^{\text{th}}$ eigenvalue $\varepsilon_{10\,000}$,
and (iii) for the lowest eigenvalue $\varepsilon^{\text{WW}}_0$ of equation~(\ref{DoubleWellPotential}),
for which $P_{\text{est}}(x)\approx{2} (x-1)^2(x+2)/{(3s\ln 10)}$.}
\end{minipage}
\end{center}

It is reassuring that the obtainable precision
can be predicted to within a few digits: Choosing a too small $x$ leads
to the solution of the wrong eigenvalue problem; choosing a too large $x$
leads to a waste of CPU cycles. The obtainable precision at fixed $x$
may be improved by some number of decimals by using a better Robin
boundary condition found by asymptotic a\-na\-lys\-is of
equation~(\ref{TypeOfEquations}). Such improvement might be useful in
situations where a few tens of decimals precision is sufficient.

\section{Method for solving the Schr{\"o}dinger equation}
\label{Method}

We have postponed the description of our numerical method of
solving equation (\ref{TypeOfEquations}), due to its (perhaps)
disappointingly naive simplicity: We make a brute force
summation of its Taylor expansion,
\begin{equation}
      \psi(x) = x^{\sigma}\,\sum^N_{m=0} a_m x^{2m},
     \label{TaylorExpansion}
\end{equation}
where $\sigma=0$ or $1$ depending on the parity of the solution.
The coefficients $a_m$, or more precisely $A_m(x)\equiv a_m\, x^{2m}$,
are generated recursively from equation (\ref{TypeOfEquations}), starting with
$A_0(x)=1$. Only the $M+1$ last coefficients need to be considered at any
time while the sum is accumulated; hence the memory requirement is low.

Since equation~(\ref{TypeOfEquations}) has no singular points in the finite plane
the sum will eventually converge {\em very fast\/}. The number of terms $N$
needed in the sum (\ref{TaylorExpansion}) can be chosen automatically by
the recursion/summation routine, but may also be {\em a apriori\/}
estimated. The calculation is done in very-high-precision floating-point
arithmetic using the CLN C++ library of numbers \cite{CLN}.

\section{Numerical loss of accuracy}
\label{AccuracyLoss}

Note that for the harmonic oscillator
the summation (\ref{TaylorExpansion}) means
computing its ground state $\psi_0(x) = \text{e}^{-x^2/2}$ for large $x$
by Taylor expansion. This is certainly not the recommended method of
computation, due to large cancellations and roundoff errors. However, by
calculating with sufficient numerical precision --- which is not
prohibitively large --- it actually works quite well. Further,
the location of eigenvalues are determined by $\varepsilon$-values
where $\psi(x)=\psi(x;\varepsilon)$ changes sign very rapidly with
$\varepsilon$.
The computation
of eigenvalues only is less sensitive to cancellations.
  
Nevertheless, the effects of roundoff and cancellations must be considered.
Computing with high-precision floating point numbers with $D$ decimals
accuracy means that the value of $\psi(x)$ in equation~(\ref{TaylorExpansion}) is
accumulated from numbers with a mantissa of $(D\ln 10/\ln 2)$ bits.
A contribution of magnitude $10^{\Delta D}$ to the sum will thus
have a round-off error of order $10^{\Delta D - D}$.
$S$ terms of the same magnitude is expected to increase this
error by a factor $\sqrt{S}$ (with symmetric roundoff), which is an
insignificant increase. In principle there might also be error amplification
in the recursion relation, but we have not observed signatures of such.
Thus, we estimate the numerical accuracy loss to $\Delta D$ decimal
digits, where
\begin{equation}
    10^{\Delta D} = \mathop{\text{max}}_m \left\{\, \vert A_m(x) \vert \,\right\}.
\end{equation}
The largest term in the sum (\ref{TaylorExpansion}) is found numerically by moni\-toring
the recursion/summation routine. It may also be
{\em a priori\/} estimated. We have done the latter in two ways:
First, by asymptotic analysis of the recursion relations
in some simple situations (those described in
sections \ref{Example1} and \ref{Example3}).
This analysis rapidly becomes complicated. Second,
all analytic and numerical results found are consistent with the assumption that
\begin{equation}
  \mathop{\text{max}}_m \left\{\, \vert A_m(x) \vert \,\right\} = 
  \mathop{\text{max}}_{\varphi} 
  \left\{\, \vert \psi(x \text{e}^{\text{i}\varphi}) \vert \,\right\},
  \label{Assumption}
\end{equation}
where $\psi(x \text{e}^{\text{i}\varphi})$ can be estimated from
a WKB-approximation. Equation (\ref{Assumption}) is based on the assumptions
that (i) there is always a point on the circle
$x\text{e}^{\text{i}\varphi}$ where cancellations are insignificant
in the sum (\ref{TaylorExpansion}),
and (ii) the main contri\-bu\-tions to the sum come from relatively
few terms around the maximum term.\footnote{Actually, the number of terms contributing to
$\mathop{\text{max}}_{\varphi} \left\{\, \vert \psi(x \text{e}^{\text{i}\varphi}) \vert \,\right\}$
is an unimportant correction to $\Delta D$.
For the example in section \ref{Example1} we used $x=152$, for which
the estimated maximum is about $10^{508\,386}$, and summed less than $10^7$
terms of the Taylor series.
Whether the maximum is contributed from one single or almost all terms of the sum
makes only a few decimals change in $\Delta D$.}
There may be parameter combinations where the
first assumption fails, in which case the equality sign
in equation~(\ref{Assumption}) should be replaced by $\ge$
(less helpful for estimations).

For the example in section \ref{Example1}
we find 
\begin{align}
  P_{\text{est}}(x)= \frac{2\,x^3}{{3\ln 10}},\quad
  \Delta D = \frac{x^3}{3\ln 10}.
\end{align}
Thus, to compute the ground state to $P$ decimals accuracy
we must evaluate the wavefunction at
$x = \left[\left(\frac{3}{2}\ln 10\right) P\right]^{1/3}$.
If we want to evaluate the wavefunction to $P$ decimals at this $x$
we must choose a numerical precision of $D=\frac{3}{2} P$ decimals,
since we will loose $\Delta D=\frac{1}{2} P$ decimals to roundoff errors.
However, if we only want to find the eigenvalue $\varepsilon$ we experience
a compensating {\em accuracy gain\/}.
An uncertainty $\delta\psi$ in the wavefunction translates to
an uncertainty 
\begin{equation}
   \delta\varepsilon = \left(\frac{\partial\psi}{\partial\varepsilon}\right)^{-1} \delta\psi
\end{equation}
in the eigenenergy. In this example
$\left(\partial\psi/\partial\varepsilon\right) \approx \text{e}^{x^3/3} \approx 10^{P/2}$.
Hence, the accuracy loss due to roundoff is completely compensated by the accuracy gain
caused by a very large $\left(\partial\psi/\partial\varepsilon\right)$. This implies
that our method of computing eigenvalues may work well even for ordinary
precisions $P$. There are many cases where such complete compensation occur.

The example in section \ref{Example3} is different. We find
\begin{align}
    P_{\text{est}}(x) &\approx \frac{2}{3s\ln 10}\,
    \left(x-1\right)^2\left(x+2\right),\nonumber\\
    \Delta D       &\approx \frac{1}{s\ln 10}\,\left(\frac{1}{3} x^3 + \frac{1}{2}x\right),\\
    \frac{\partial\psi}{\partial\varepsilon} &\approx 10^{P_{\text{est}}(x)/2}.\nonumber
\end{align}
In this case the accuracy loss is not fully compensated. We note that
$\Delta D \approx 5/(6s\ln 10)$ (large when $s$ is small)
when $P$ is chosen small. In this case it will be computationally
very expensive to obtain results to a few decimals of accuracy by this method.
However, for very large $P$ the computational cost is similar to
the example in section \ref{Example1}.
The example in section \ref{Example2} is similar to the one 
section \ref{Example3}, only with more cumbersome expressions.

\section{Numerical observations}
\label{NumericalObservations}

\hspace{-3.0em}
\begingroup
  \makeatletter
  \providecommand\color[2][]{%
    \GenericError{(gnuplot) \space\space\space\@spaces}{%
      Package color not loaded in conjunction with
      terminal option `colourtext'%
    }{See the gnuplot documentation for explanation.%
    }{Either use 'blacktext' in gnuplot or load the package
      color.sty in LaTeX.}%
    \renewcommand\color[2][]{}%
  }%
  \providecommand\includegraphics[2][]{%
    \GenericError{(gnuplot) \space\space\space\@spaces}{%
      Package graphicx or graphics not loaded%
    }{See the gnuplot documentation for explanation.%
    }{The gnuplot epslatex terminal needs graphicx.sty or graphics.sty.}%
    \renewcommand\includegraphics[2][]{}%
  }%
  \providecommand\rotatebox[2]{#2}%
  \@ifundefined{ifGPcolor}{%
    \newif\ifGPcolor
    \GPcolorfalse
  }{}%
  \@ifundefined{ifGPblacktext}{%
    \newif\ifGPblacktext
    \GPblacktexttrue
  }{}%
  \let\gplgaddtomacro\g@addto@macro
  \gdef\gplbacktext{}%
  \gdef\gplfronttext{}%
  \makeatother
  \ifGPblacktext
    \def\colorrgb#1{}%
    \def\colorgray#1{}%
  \else
    \ifGPcolor
      \def\colorrgb#1{\color[rgb]{#1}}%
      \def\colorgray#1{\color[gray]{#1}}%
      \expandafter\def\csname LTw\endcsname{\color{white}}%
      \expandafter\def\csname LTb\endcsname{\color{black}}%
      \expandafter\def\csname LTa\endcsname{\color{black}}%
      \expandafter\def\csname LT0\endcsname{\color[rgb]{1,0,0}}%
      \expandafter\def\csname LT1\endcsname{\color[rgb]{0,1,0}}%
      \expandafter\def\csname LT2\endcsname{\color[rgb]{0,0,1}}%
      \expandafter\def\csname LT3\endcsname{\color[rgb]{1,0,1}}%
      \expandafter\def\csname LT4\endcsname{\color[rgb]{0,1,1}}%
      \expandafter\def\csname LT5\endcsname{\color[rgb]{1,1,0}}%
      \expandafter\def\csname LT6\endcsname{\color[rgb]{0,0,0}}%
      \expandafter\def\csname LT7\endcsname{\color[rgb]{1,0.3,0}}%
      \expandafter\def\csname LT8\endcsname{\color[rgb]{0.5,0.5,0.5}}%
    \else
      \def\colorrgb#1{\color{black}}%
      \def\colorgray#1{\color[gray]{#1}}%
      \expandafter\def\csname LTw\endcsname{\color{white}}%
      \expandafter\def\csname LTb\endcsname{\color{black}}%
      \expandafter\def\csname LTa\endcsname{\color{black}}%
      \expandafter\def\csname LT0\endcsname{\color{black}}%
      \expandafter\def\csname LT1\endcsname{\color{black}}%
      \expandafter\def\csname LT2\endcsname{\color{black}}%
      \expandafter\def\csname LT3\endcsname{\color{black}}%
      \expandafter\def\csname LT4\endcsname{\color{black}}%
      \expandafter\def\csname LT5\endcsname{\color{black}}%
      \expandafter\def\csname LT6\endcsname{\color{black}}%
      \expandafter\def\csname LT7\endcsname{\color{black}}%
      \expandafter\def\csname LT8\endcsname{\color{black}}%
    \fi
  \fi
  \setlength{\unitlength}{0.0500bp}%
  \begin{picture}(5040.00,3528.00)%
    \gplgaddtomacro\gplbacktext{%
      \csname LTb\endcsname%
      \put(1254,440){\makebox(0,0)[r]{\strut{}$10$}}%
      \put(1254,857){\makebox(0,0)[r]{\strut{}}}%
      \put(1254,1273){\makebox(0,0)[r]{\strut{}$10^3$}}%
      \put(1254,1690){\makebox(0,0)[r]{\strut{}}}%
      \put(1254,2107){\makebox(0,0)[r]{\strut{}$10^5$}}%
      \put(1254,2523){\makebox(0,0)[r]{\strut{}}}%
      \put(1254,2940){\makebox(0,0)[r]{\strut{}$10^7$}}%
      \put(1765,220){\makebox(0,0){\strut{}$10^3$}}%
      \put(3022,220){\makebox(0,0){\strut{}$10^4$}}%
      \put(4279,220){\makebox(0,0){\strut{}$10^5$}}%
      \put(4632,220){\makebox(0,0){\strut{}$P$}}%
    }%
    \gplgaddtomacro\gplfronttext{%
      \csname LTb\endcsname%
      \put(3723,1053){\makebox(0,0)[r]{\strut{}$\Delta D$}}%
      \csname LTb\endcsname%
      \put(3723,833){\makebox(0,0)[r]{\strut{}$N$}}%
      \csname LTb\endcsname%
      \put(3723,613){\makebox(0,0)[r]{\strut{}$T\; [\text{ms}]$}}%
    }%
    \gplbacktext
    \put(0,0){\includegraphics{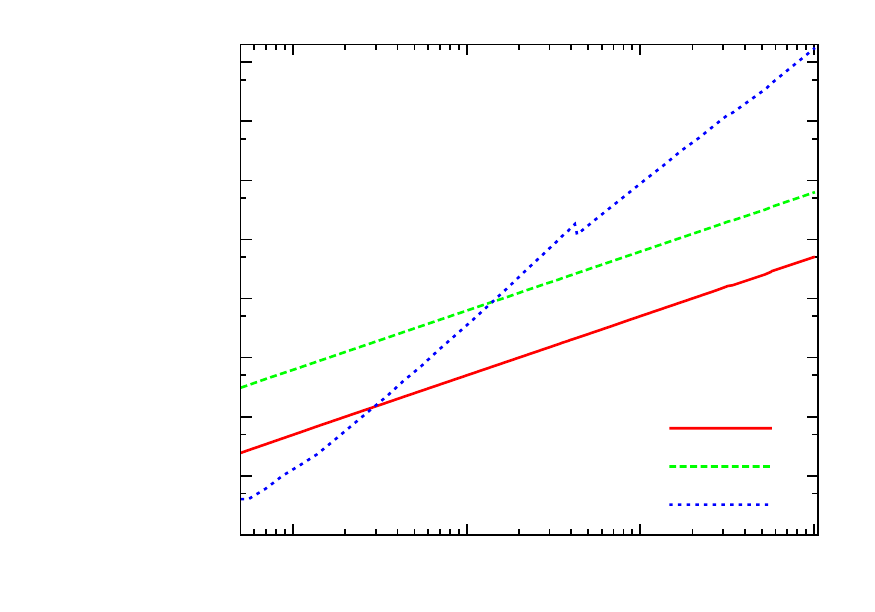}}%
    \gplfronttext
  \end{picture}%
\endgroup

\vspace{-2ex}
\begin{center}
\begin{minipage}{0.9\linewidth}{ {\bf Figure 2:} Some observed
behaviour when calculating the ground state energy of
equation~(\ref{Anharmonic_oscillator})
to $P$ decimals precision:
(i) The accuracy loss $\Delta D$ (in decimals) due to roundoff error; in this case $\Delta D\sim P/2$.
(ii) The number $N$ of terms needed in the Taylor expansion~(\ref{TaylorExpansion}),
proportional to $P$.
(iii) The time $T$ used to evaluate one wavefunction to required precision, 
locally we find $T \sim P^\nu$,
where $\nu\approx 2.6$ around $P=10\,000$ and $\nu\approx2.13$ around $P=200\,000$.
(Datapoints for $P>300\,000$ are not generated under identical conditions.)
}
\end{minipage}
\end{center}

We monitor many parameters during the numerical
computations. A sample of those are shown in figure 2,
from computation of the ground state energy of
equation (\ref{Anharmonic_oscillator}). We plot the
numerically observed accuracy loss $\Delta D$; this
agrees with the {\em a priori\/} estimate. This is also
observed for all other investigated cases.

Further, the number $N$ of terms required in the
sum (\ref{TaylorExpansion}) seems to grow linearly with
precision $P$. This is also the case for other examples,
only with different coefficients of proportionality.

The total time to make one evaluation of the
wavefunction $\psi(x)$ also behaves as expected
from the number of terms in the sum, and the
time needed to multiply two high-precision
numbers. The drop in computation time
near $P=42\,000$ does not seem to be an
artifact of variations in the computational
environment. We believe it is due to a change of
multiplication algorithm at this precision.

\section{Possible extensions}
\label{PossibleExtensions}

The possibilites of extending our method to
other systems are somewhat limited. It seems
straightforward to generalize to non-symmetric 
one-dimensional potentials, to Schr{\"o}dinger
equations which have only one regular singular point in the
finite plane, and to small systems of such
equations. We believe that systems
with two (unseparable) degrees of freedom can
be constructively approached.

The evaluation of unnormalized wavefunctions
is certainly possible, with a time requirement
proportional to the number of evaluation
points. We do not rule out the possibility
of computing normalized wave functions to
high precision.

Another interesting extension
is towards very-high-pre\-ci\-sion computation
of Green functions for the same class of models.

\section{Acknowledgement}
\label{acknowledgement}

This work was supported in part by the Higher Education Commision of Pakistan (HEC)
and the ``SM{\AA}FORSK'' program of the Research Council of Norway. KO thanks
NORDITA, Stockholm for hospitality and support July-December 2009, where parts of this work
was performed.





\begin{thebibliography}{00}


\bibitem{BenderWu}
  C.M.~Bender and T.T.~Wu,
  {\em Large-Order Behaviour of Perturbation Theory\/},
  Physical Review Letters {\bf 27}, 461 (1971)

\bibitem{BenderWuII}
   C.M.~Bender and  T.T.~Wu,
   {\em Anharmonic Oscillator. II. A Study of Perturbation Theory in Large Order\/},
   Physical Review {\bf D7}, 1620 (1972)

\bibitem{ZinnJustin}
   J.~Zinn-Justin,
   {\em Expansion around instantons in quantum mechanics\/},
   J.~Math Phys. {\bf 22}, 511 (1981)






\bibitem{Bender_etal}
   C.M.~Bender, K.~Olaussen and P.S.~Wang,
   {\em Numerological analysis of the WKB approximation in large order\/},
   Physical Review {\bf D16}, 1740 (1977)

\bibitem{CLN} B.~Haible and R.B.~Kreckel,
   {\em CLN -- Class Library for Numbers\/},
   {http://www.ginac.de/CLN/}

\bibitem{GMP} T.~Granlund and collaborators,
   {\em GMP -- The GNU Multiple Precision Arithmetic Library\/},
   {http://gmplib.org/}

\bibitem{AAKI} A.~Mushtaq, A.~Noreen, K.~Olaussen, and I.~{\O}verb{\o},
  In preparation.

\end{thebibliography}



\end{document}